\documentclass[showpacs,prl,amsmath,amssymb]{revtex4}
\usepackage{graphicx}
\usepackage{amsmath}
\usepackage{amssymb}
\usepackage{units}

\begin{document}
\title{A scaling theory for the quasi-deterministic limit}
 \author{David A. Kessler}
 \email{kessler@dave.ph.biu.ac.il}
\affiliation{Department of Physics, Bar-Ilan University,
 Ramat-Gan 52900 Israel }
\author{Nadav M. Shnerb}
\email{shnerbn@mail.biu.ac.il}
\affiliation{Department of Physics, Bar-Ilan University,
 Ramat-Gan 52900 Israel }

\begin{abstract}

Deterministic rate equations are widely used in the study of
stochastic, interacting particles systems.  This approach assumes
that the  inherent noise, associated with the discreteness of the
elementary constituents, may be neglected when the number of
particles $N$ is large. Accordingly, it fails close to the
extinction transition, when the amplitude of stochastic fluctuations
is comparable with the size of the population. Here we present a
general scaling theory of the transition regime for spatially
extended systems. Two fundamental models for out-of-equilibrium
phase transitions  are considered: the
Susceptible-Infected-Susceptible (SIS) that belongs to the directed
percolation equivalence class, and the
Susceptible-Infected-Recovered (SIR) model belonging to the dynamic
percolation class. Implementing the Ginzburg criteria we show that
the width of the fluctuation-dominated  region scales like
$N^{-\kappa}$, where $N$ is the number of individuals per site and
$\kappa = 2/(d_u-d)$,  $d_u$ is the upper critical dimension. Other
exponents that control the approach to the deterministic limit are
shown to depend on $\kappa$.  The theory is extended to include the
corrections to the front velocity above the transition. It is
supported by the results of extensive numerical simulations for
systems of various dimensionalities.
\end{abstract}


\maketitle

\section{introduction}

The connection between a stochastic model of particle reactions (or
equivalently, birth-death processes) and its associated
deterministic rate equations is a topic of continuing interest.  The
common intuition is that the rate equations are not only
qualitatively correct, but indeed provide,  when the number of
interacting particles is large, a quantitatively accurate
approximation.  This intuition is given concrete support by the
$\Omega$ expansion of van Kampen~\cite{vanKampen}.  However, there
are a number of situations in which this picture is too naive and
needs to be refined.  One by now classic example of this is the
exponentially small rate of extinction for a system with an
absorbing state~\cite{extinct}, which dominates the long-time
dynamics, and is completely missed by the rate equations.  Another
example is the anomalously large corrections~\cite{derrida,nks} to
the front velocity in stochastic systems which exhibit propagation
into an unstable state; e.g., systems whose rate equation is the
Fisher-Kolmogorov equation.

A system which captures features of both these examples is the
spatially extended version~\cite{KS0} of the classic SIS
(Susceptible-Infected-Susceptible) infection model of Weiss and
Dishon~\cite{sis}.  In this model, contact (either on-site or
nearest-neighbor) between an infected individual and a susceptible
can, with some probability, convert the susceptible into an
infected.  Infected individuals spontaneously leave the infected
state, reverting to susceptible. The well-mixed SIS system for
sufficiently high infection probability possesses an endemic state,
with a essentially constant level of infecteds, which however is
subject to an exponentially small rate of extinction due to the
existence of the absorbing state of zero infecteds. In addition, the
deterministic rate equation is of Fisher-Kolmogorov type, and so a localized infection in the non-well-mixed case exhibits at
the deterministic level  an infection wave which propagates
at constant velocity.  However, the stochastic system exhibits not a
bifurcation but rather a phase transition, characterized by the
anomalous (for dimension $D<4$) scaling exponents of the
directed-percolation (DP) problem \cite{hinrichsen}.

The connection between  this complicated statistical behavior
and the deterministic rate equations, which should be valid in the
large-$N$ limit ($N$ being the total number of individuals, both
susceptible and infected, on each site), is thus a natural topic for
investigation, a study we initiated in a recent paper~\cite{KS0}
(I).  There we found that in one spatial dimension,  the large-$N$
behavior was governed by a scaling law with an exponent which we
called $\kappa \approx 0.66$.  For example, the phase transition
point was shifted from the deterministic bifurcation point by an
amount proportional to $N^{-\kappa}$.  Investigating the correlation
length, $\xi$, we found that there was a scaling collapse so that
$\xi N^{\tau}$, with $\tau= 0.41$ was a function of $N^{\kappa}$
times the distance to the deterministic bifurcation point.

In this paper, we show that this behavior is in fact a
nonequilibrium version of what one may call the Ginzburg crossover.
A fundamental concept in equilibrium field theory is that of the
Ginzburg criterion, which states under which circumstances the noise
is relevant.  This is of course what predicts the existence of an
upper critical dimension (UCD), above which the noise does not
affect the long distance behavior and so the scaling is mean-field
like. The logic underlying the Ginzburg criterion implies that if
one could ``dial" down the noise, the system would look more and
more classical, and a crossover (which we will call the Ginzburg
crossover) between the classical and noise-dominated regimes should
become apparent, with the noise dominating the very long distance
behavior (below the UCD, of course).  This program has been
implemented in the context of the equilibrium finite-range Ising
model, where spins interact with all their neighbors out to a
distance $R$.  As $R$ increases, each spin is interacting with what
is more and more closely approximating the mean-field, and
mean-field behavior at short scales sets in.  The problem can be
carried out analytically for the finite-range spherical
model~\cite{pelissetto}, and has been investigated via simulation in
the finite-range Ising model~\cite{binder}.

Another fundamental epidemics model considered in I is
the SIR infection model of Kermack and McKendrick~\cite{sir}, where
recovered individuals are immune to further infection.  The critical
behavior of this model is governed by the dynamic percolation
exponents, with an upper critical dimension of 6. In I
we have carried a numerical investigation of the SIR model also,
showing that the theory converges, in the large $N$ limit, to its
mean-field limit with  scaling exponents $\kappa$ and $\tau$ that
differ from those of the SIS (DP) model. Here we consider again the
SIR model and derive analytically the relevant exponents using
the same theory  of the Ginzburg crossover, applied to the different universality class of the SIR model.

The main quantity used in the theory of epidemics to characterize
the transmission potential of a disease is the basic reproduction
rate $R_0$, which is the expected number of secondary cases produced
by a primary case in a population that is completely susceptible
\cite{anderson}. In the absence of demographic noise (e.g., in an
infinite-dimensional model) the transition takes place at $R_0 =1$.
Noise shifts the transition to higher values of $R_0$, but, as we
will show below, the value $R_0=1$  still admits a special feature:
the renormalized distance between $R_0 =1$ and the actual transition
point is $N$-independent as long as $\kappa < 1$. This interesting
feature allows one to examine the scaling properties numerically in
a very efficient manner, as it saves the effort needed to identify
the location of the transition point for each $N$ separately. This
feature is utilized here when we compare the expected results with
numerical simulations.

\section{The well mixed SIS dynamics and the transition zone}

We  first review the well-mixed version of the SIS model, and the
relation of the stochastic model to the deterministic equations that
determine the evolution of the system. Although this stochastic model has
already been analytically solved,  the discussion allows us to present the
concepts that we intend to  use below and to set the mathematical
framework used in the study  of the spatial models.

Let us consider a population of exactly $N$ individuals, some of
them are infected (I) and the rest are susceptible ($S = N-I$). The
allowed processes are infection (with rate $\alpha/N$, this is the
type II model of \cite{anderson}) and recovery (with rate $\beta$):

\begin{equation}
S+I \stackrel{\alpha/N}{\longrightarrow} 2I  \qquad I
\stackrel{\beta}{\longrightarrow} S.
\end{equation}
The corresponding master equation for the microscopic process can be
formulated in terms of $P_n$, the chance to have  $n$ infected
individuals:
\begin{equation} \label{master0}
\dot{P_n} = \beta \left( -nP_n + (n+1)P_{n+1} \right) +
\frac{\alpha}{N} \left[-n(N-n)P_n + (n-1)(N-(n-1))P_{n-1} \right].
\end{equation}
Defining $\langle I \rangle= \sum_n nP_n$ as the expected number of
infecteds, one finds after index rearrangement:
\begin{equation} \label{eq3}
\langle \dot{I} \rangle = -\beta \langle I \rangle +
\frac{\alpha}{N} \sum_n n(N-n) P_n = (\alpha - \beta) \langle I
\rangle - \frac{\alpha}{N} \langle I^2 \rangle.
\end{equation}
The essence of the van Kampen $\Omega$ expansion is that this
equation closes if $\langle I \rangle \gg 1$, so that the variance
of $I$ makes a negligible contribution, giving the standard logistic
equation
\begin{equation}
\langle \dot{I} \rangle =  (\alpha - \beta) \langle I \rangle -
\frac{\alpha}{N} \langle I \rangle^2. \label{mf1}
\end{equation}
Since $0\le \langle I \rangle \le N$, it is necessary for $N$ to be
large, in order for the rate equation, Eq. (\ref{mf1}), to be valid.  This, however, is not sufficient.  If $\alpha
> \beta$,  Eq. (\ref{mf1})
 has an attractive fixed point at $\langle I \rangle = I_0  = N (1-1/R_0)$, where
$R_0 \equiv \alpha/\beta$ is the primary reproductive number and  $I_0$
is indeed large if  $N$ is large, as required.
Although the system admits an absorbing state at $I=0$, the chance
of a giant fluctuation that takes the system from $I_0$ to zero is
exponentially small in $I_0$, thus when $N \to \infty$ stochastic
extinction (fadeout) is impossible once the system reaches its
steady state. However, if the number of infected individuals in the
initial state is small, stochastic effects are transiently present
even in the $N\to \infty$ limit. For example, introducing one
infected individual results in either short-time extinction (with
probability $1/R_0$) or an endemic state (with probability
$(R_0-1)/R_0)$.  If $R_0 = 1$ exactly, at the $N \to \infty$ limit
the system performs an unbiased random walk in $n$, the number of
infecteds, and the theory of first passage times tells us that the
chance of extinction is still unity, but the probability $P(q)$ to
have $q$ infection events scales like $q^{-3/2}$.

At finite $N$ the situation is more complex. Now the steady state of
Eq. (\ref{mf1}) corresponds to a finite number of infected
individuals in the endemic state, which mean that a finite, but
large, fluctuation may cause a fadeout. The chance for such a
fadeout is large when $R_0$ is close to one, i.e., when the
attractive fixed point corresponds to only a few individuals.
Instead of having a sharp extinction to proliferation transition at
$R_0 = 1$,  now the transition is ``soft": defining ${\widetilde\Delta}
= R_0 -1$ as the distance from the transition, $I_0 \sim N {\tilde
\Delta}$; a metastable state exists only if this quantity (the
distance of the stable solution from the absorbing state) is larger
than the typical fluctuation size, $\sqrt{N}$, thus a transition
\emph{zone} of width ${\widetilde\Delta} \sim N^{-1/2}$ occurs between
the extinction and the proliferation regimes. As shown in
\cite{kess}, $P(q)$ decays exponentially in the extinction phase
${\widetilde\Delta} <0$, has a peak at $exp{(const\cdot N)}$ at the
endemic phase ${\widetilde\Delta} \gg 1/\sqrt{N}$, and decays like
$q^{-3/2}$ with a cutoff at $N$ in the transition zone. Note that
the width of the transition zone goes to zero as $N$ approaches
infinity, recovering the sharp transition at ${\widetilde\Delta}=0$
that characterizes the deterministic theory.

\section{The absence of self-interaction}

The derivation of Eq. (\ref{mf1})  from Eq. (\ref{eq3}) involves the
neglect of ${\cal O} (1/N)$ terms. In particular one can easily
see that the rate of infection when only one infected individual
appears in a population of size $N$ is $\alpha (1-1/N)$, so the
transition occurs at $R_0 = 1 + 1/N$. This result reflects the most
trivial effect of discretization, namely, the absence of self-interactions \cite{durrett}: an infected individual cannot infect
itself, so the effective size of the population "seen" by the first
infected is $N-1$ instead of $N$.  There are presumably other nonsingular $1/N$ corrections
to the transition point, but for convenience we will refer to all these $1/N$ corrections as the ``self-interaction" effect.

Putting this fact together with the discussion of the last section,
we realize that there are two $N$ dependent functions that control
the transition: one is the  ${\cal O}( 1/N)$ shift of the transition
point, the other is the width of the "quantum" regime (the region
above the transition point in which the system is controlled by
demographic fluctuations) that scales, in the well-mixed limit, like
$N^{-1/2}$. As $N \to \infty$ the shift is negligible with respect
to the width of the transition zone, so there is only one scale in
the problem, $\Delta \sim N^{-1/2}$. However this behavior is not
generic. As we will show below, in some cases the width of the
transition zone is much \emph{narrower} than $1/N$, and in these
cases one should take into account the two scales.

\section{Spatial  SIS model and the transition zone}

What happens if the system is extended? For the sake of concreteness
let us focus on the example of an infinite one dimensional array of
patches with $N$ individuals on each patch. The probability per unit
time of a susceptible on the $n$th site being infected by a given
sick agent residing at this site is $\alpha (1-\chi)/N$ and  of
being infected by a given infected resident of one of the
neighboring sites is $\alpha \chi/2N$ (in a $d$ dimensional system,
this chance will be $\alpha \chi/Nd$). This corresponds to the
"travelers model" considered in Ref. \cite{yosibz}. The
deterministic rate, or mean-field (MF), equations, are
\begin{eqnarray}
\dot{I_n} &=& -\beta I_n + \frac{\alpha (1-\chi)}{N} I_n (N-I_n)
+\frac{\alpha \chi}{2N} (N-I_n)(I_{n+1} + I_{n-1}) \nonumber \\
&=&   \frac{\alpha \chi}{2} \nabla^2 I + (\alpha - \beta)I_n -
\frac{\alpha}{N} I_n^2 + \frac{\alpha \chi}{2N} I \nabla^2 I
\label{mf2}\end{eqnarray}
where $\nabla^2$ stands for the discrete
version of the  Laplacian operator. The last, nonlinear diffusion,
term,  does not materially affect the dynamics (naive dimensional
analysis shows that it is an irrelevant operator). Without this term
one recognizes,  on the MF level, the celebrated Fisher (or FKPP
\cite{fisher}) equation for invasion of a stable into an unstable
phase, with a sharp transition at $\alpha = \beta$ (or $R_0 \equiv
\alpha / \beta = 1$), and front propagation with a velocity of
$2\sqrt{\alpha \chi \beta \widetilde\Delta/2}$, since
 the  effective diffusion constant is $\alpha
\chi/2$ and the net growth rate is $\alpha-\beta=\beta \widetilde\Delta$.

What happen when stochasticity is taken into account? If $N=1$,
i.e., there is only one agent on any site and so all infections are
nearest-neighbor (thus it is reasonable to take $\chi=1$), the
stochastic process is known as the contact process, which undergoes
a continuous phase transition from extinction to proliferation. The
``effective" infection rate is smaller than $\alpha$, since a sick
agent cannot infect its neighbor if it is already sick. The
transition happens at some $R_c >1$, e.g., here for $N=1$, $R_c
\approx 3.297$. While the exact value of $R_c$ is of course
non-universal, the extinction transition, which belongs to the
directed percolation equivalence class \cite{hinrichsen}, admits
three universal
 critical exponents:
 \begin{enumerate}
   \item The spatial correlation length diverges as
   $|\Delta|^{-\nu_{\perp}}$, where we introduce $\Delta\equiv R_0-R_c$ as the distance from the
  stochastic transition, as opposed to $\widetilde\Delta$, which measures the distance to the {\em mean-field} transition;  in 1d, $\nu_{\perp} \approx 1.09$
   \item The temporal correlation length diverges like
   $|\Delta|^{-\nu_{\parallel}}$; in 1d, $\nu_{\parallel} \approx 1.73$
   \item Above the transition the steady state density of infecteds,  $I_0$, grows
   like $\Delta^{\beta}$; in 1d, $\beta \approx 0.28$
 \end{enumerate}
The values of these critical exponents depend only on the
dimensionality of the system and not on the microscopic details of
the process. Above the critical dimension $d=4$ the exponents take
their MF values, $\nu_\perp =1/2$, $\nu_\parallel = 1$, $\beta = 1$.

As  $N$  (the number of agents on a site) increases, demographic
fluctuations become smaller. In the infinite $N$ limit one recovers
the MF transition described in Eq. (\ref{mf2}). First, the
transition point moves back to $R_c =1$; second, the values of the
critical exponent in this deterministic limit are equal to their MF
values. For example it is clear from Eq. (\ref{mf2}) that above the
transition the density scales linearly with $\Delta$, i.e., that $\beta=1$.
Below the transition $I$ is small and the nonlinear term in
Eq. (\ref{mf2}) is negligible, hence if $I(x,0) = \delta(x)$, $I(x,t) \sim
exp(-x^2/2Dt - \Delta t)$. The maximal density at $x$ occurs when $t
\sim x/\sqrt{\Delta D}$; thus the spatial profile of {\em total} infections is proportional to
$exp(-x/\xi_{\perp})$ with $\xi_\perp \sim 1/\sqrt{\Delta}$, so that
$\nu_\perp^{MF} = 1/2$.

At any finite $N$, though, close enough to $R_c$ the system is
controlled by stochastic effects, as implied by universality. As $N$
becomes large, the effects of stochasticity are restricted to a narrow
region close to the transition point, which defines the width of
the transition (``quantum") zone.

In I, we have shown numerically that close to the
transition point the spatial correlation length is given by:
\begin{equation} \label{scaling}
\xi_\perp  = A N^{-\tau} (R_{c} - R_0)^{-\nu_\perp}
\end{equation}
where the transition takes place at  $R_{c} =  1 + B N^{-\kappa}$.
The values $\kappa \approx 0.66$ and $\tau \approx 0.41$ have been
obtained numerically for different microscopic models that belongs
to the DP equivalence class and seem to be identical for the
different models up to the accuracy of the numerics. As long as
$\kappa <1$, the region in the parameter space in which the system
is controlled by stochasticity coincides with the interval between
the stochastic and the deterministic critical points; i.e., it also
scales like $N^{-\kappa}$. Rescaling appropriately the  correlation
length and the distance from the transition, our numerics (see I)
showed an whole scaling regime described by the function:
\begin{equation} \label{scaling1}
N^{\kappa - \tau/\nu_\perp} \xi ^{-1/\nu_\perp} = {\cal F} (\widetilde{\Delta}
N^\kappa)
\end{equation}
The scaling
function $\cal{F}$ vanishes linearly at a positive value of its
argument, which marks the transition point.
Notice that what enters here is $\widetilde\Delta\equiv R_0-1$, so that the behavior at the classical transition point is controlled by the
fluctuations, even though it is outside the range of the linear behavior of ${\cal{F}}$. We will see later that the story is more complicated for $\kappa>1$.

As  discussed in the introduction, this scaling behavior is the
result of a crossover between the deterministic theory and the
critical theory as the critical point is neared. We will now use
this  to derive a scaling relation between $\tau$ and $\kappa$. Then
we will obtain the value of $\kappa$  by calculating the Ginzburg
criterion for the model.

To connect $\tau$ to $\kappa$, one observes that the scaling function
${\cal F}(x)$ takes us from the stochastic regime at finite $x$
(close to the transition) to the the deterministic regime at large
 negative $x$,  corresponding to the region
 deeply below the transition. Even for $|x|$ large, the system may
 still be arbitrary close to the transition ($\widetilde\Delta$ may be
 arbitrarily small) as long as $N$ is large enough. This implies
 that in the $x \to -\infty$ limit, the correlation length must
 diverge like $\widetilde\Delta^{-1/2}$, \emph{independent of $N$}. As a
 result the leading behavior of  ${\cal F}(x)$ at large negative $x$ must obey
${\cal F}(x) \sim x^{1/2\nu_\perp}$. To cancel the $N$ dependence in
the expression
\begin{equation} N^{\kappa - \tau/\nu_\perp} \xi^{-1/\nu_\perp} =
\Delta^{1/2\nu_\perp} N^{\kappa/2\nu_\perp}
\end{equation} one must have the scaling relation
\begin{equation}
\tau = \kappa \left(\nu_\perp - \frac{1}{2}\right).
\end{equation}
Given that  we found $\kappa\approx 0.66$, this implies a value of $\tau\approx .40$,
consistent with our numerical findings.  This scaling relation also implies that we can rewrite Eq. (\ref{scaling1})
as
\begin{equation}
\xi=N^{\kappa/2} \left[{\cal{F}}\left(\widetilde\Delta N^\kappa\right)\right]^{-\nu_\perp}
\end{equation}

A similar argument is applicable to any of the quantities that
diverge at the transition. One example that will be used below is
the overall "mass" $M_N$ of a cluster, namely the average total
number of infection events before extinction. Utilizing  the same
scaling analysis, and the known mean field dependence $M_N =
1/\widetilde\Delta$, one expects that for $\kappa<1$,
\begin{equation}\label{zzz}
M_N \sim  N^{\kappa} \left[{\cal G} \left(\widetilde\Delta N^{\kappa} \right)\right]^{-\gamma}
\end{equation}
where ${\cal{G}}$ vanishes linearly at the transition point, and $\phi_M$ is the critical scaling exponent for the mass,
\begin{equation}
\gamma \approx 1.24
\end{equation}
Eq. (\ref{zzz} ) is a  useful relation that  allows us to recover
$\kappa$ directly from numerical simulations at fixed $R_0$. To
demonstrate the critical exponents one has to locate exactly  the
transition point for any value of $N$; this is indeed a very tedious
task. Instead, we can choose to simulate exactly at $R_0=1$, which
in our case implies $R_0=1$. At this point the argument of the
scaling function is exactly zero, independent of $N$, so the mass
scales like $N^\kappa$. A plot of  $M_N/N^{\kappa}$  vs. $N$  at
$R_0=1$ must converge to a constant in the large $N$ limit. Below we
will test this condition to verify numerically the predictions
of our theory for $\kappa$, as explained in the next section.

However, this strategy works only for $\kappa < 1$. As explained
above, for higher values of $\kappa$ the trivial self-interaction
shift of the transition point is not negligible in the $N \to
\infty$ limit. Thus, as will be exemplified below, for dimensions
where $\kappa > 1$ one has to find first the transition point at
$R_c = 1 + {\cal O}(1/N)$, and only near that point the transition
region manifests itself.

\section{The exponent $\kappa$ and the  Ginzburg crossover}

Determining $\kappa$, thus, is enough to know everything about the
quasi-deterministic regime. To find  the value of $\kappa$ we adopt
here a Ginzburg criterion approach,
 looking for the leading perturbative correction in
inverse powers of $N$, and associate the stochastic regime with the
region where this leading correction is ${\cal O}(1)$.

As a platform for the perturbative analysis we have chosen the
Peliti-Doi field theoretic technique \cite{pelity} (see \cite{cardy}
for details). Starting with  the master equation for the SIS
process, at a single site (zero dimensional system) with $N$
individuals presented above.  Eq. (\ref{master0}) may be written as
\begin{equation}
 \dot {\psi} = - {\cal H}\psi
 \end{equation} where
 \begin{equation}
 \psi \equiv \sum_n P_n |n\rangle .
 \end{equation}
   Using the
creation-annihilation operators $a |n\rangle = n |n-1\rangle$ and
$a^\dag |n\rangle = |n+1\rangle$, the ``Hamiltonian" takes the form
\begin{equation}
{\cal H}/\beta =  (a^\dag a - a) + R_0 (a^\dag a - a^\dag a^\dag a)
+ \frac{R_0}{N} (a^\dag -1)(a^\dag a a^\dag a).
\end{equation}
Using the commutation relation $[a^\dag,a] = 1$ and shifting from
$a^\dag$ (that have a vacuum expectation value of unity
\cite{cardy}) to $\bar{a} = a^\dag -1$ one obtains:
\begin{equation}
{\cal H}/\beta = (1 - R_0(1-\frac{1}{N})) \bar{a} a - R_0
(1-\frac{1}{N}) \bar{a} \bar{a} a - \frac{R_0}{N} \left( \bar{a}
\bar{a}\bar{a}aa+2\bar{a}\bar{a}aa+\bar{a} aa \right).
\end{equation}
The first,  ``mass" term of the Hamiltonian determines the
transition point: the system is in the active phase when the
(renormalized) mass becomes negative. If $N \to \infty$, an outbreak
may occur at $\alpha
>\beta$, i.e., the transition happens when $\widetilde\Delta = R_0-1 = 0$. The $1/N$
correction to this result reflects, again,  the absence of
self-interactions.

 Formally, the time
evolution of $\psi$ is given by:
\begin{equation}
\psi(t) = e^{-Ht} \psi(t=0).
\end{equation}
With the aid of time slicing and the coherent state representation
one may arrive at a path integral representation of the evolution in
time where the former creation-annihilation operators are replaced
by complex-valued fields defined over a continuous space-time
\cite{cardy}:
\begin{equation}
\psi(t) = \int {\cal D}a {\cal D}\bar{a}\, e^{-{\cal S}_0(\bar{a},a) - {\cal S}_1 (\bar{a},a)} \psi(0)
\end{equation}
 where
\begin{equation}\label{action}
{\cal S}_0 = \int d^dx\, dt\ \bar{a}(\vec{x},t) [ \partial_t - D
\nabla^2  - m] a(\vec{x},t) .
\end{equation}
with $m = {\widetilde\Delta} - \frac{R_0}{N}$ and
\begin{equation}\label{S1}
{\cal S}_1 = R_0 \int d^dx\, dt\  \left[(1-\frac{1}{N}) \bar{a} \bar{a} a -
\frac{1}{N} \left( \bar{a}
\bar{a}\bar{a}aa+2\bar{a}\bar{a}aa+\bar{a} aa \right)\right] .
\end{equation}
The renormalized  values for all the constants in the problem may be
obtained perturbatively by averaging over the cumulant  expansion of
$exp(-{\cal S}_1)$ with weight $exp(-{\cal S}_0)$. The free
propagator, in terms of spatial Fourier components, is
\begin{equation} \label{prop} \langle \bar{a} (k',t') a(-k,t) \rangle =
\delta(k'+k) \theta(t-t') e^{(-k^2 + m)(t-t`)}.
\end{equation}

Here we are not really interested in the exact values of the
perturbative corrections. All we are looking for is the width of the
transition zone in the limit $\widetilde\Delta \to 0$ and $N \to \infty$. If a
perturbative correction is proportional to $N^{-y}\widetilde \Delta^{-x}$,
this correction becomes important (i.e., of order unity) when $\widetilde\Delta =
N^{-y/x}$. There are many possible perturbative corrections with
different $x$ and $y$, but $\kappa$ is determined by the one that
corresponds to the minimal value of $y/x$. In appendix 1 we will
analyze the various elements of the perturbative expansion and conclude that
\begin{equation}\label{siskappa}
\kappa = \frac{2}{4-d} ;
\end{equation}
in particular $\kappa$ is $2/3$ in one dimension, with almost perfect
agreement with the numerical results reported in I. Moreover
our result for a well mixed system (zero dimensions) is indeed
$\kappa = 1/2$, again with perfect agreement with the known results
in that case.

\begin{figure}
\vspace*{-.4in}\includegraphics[width=0.7\textwidth]{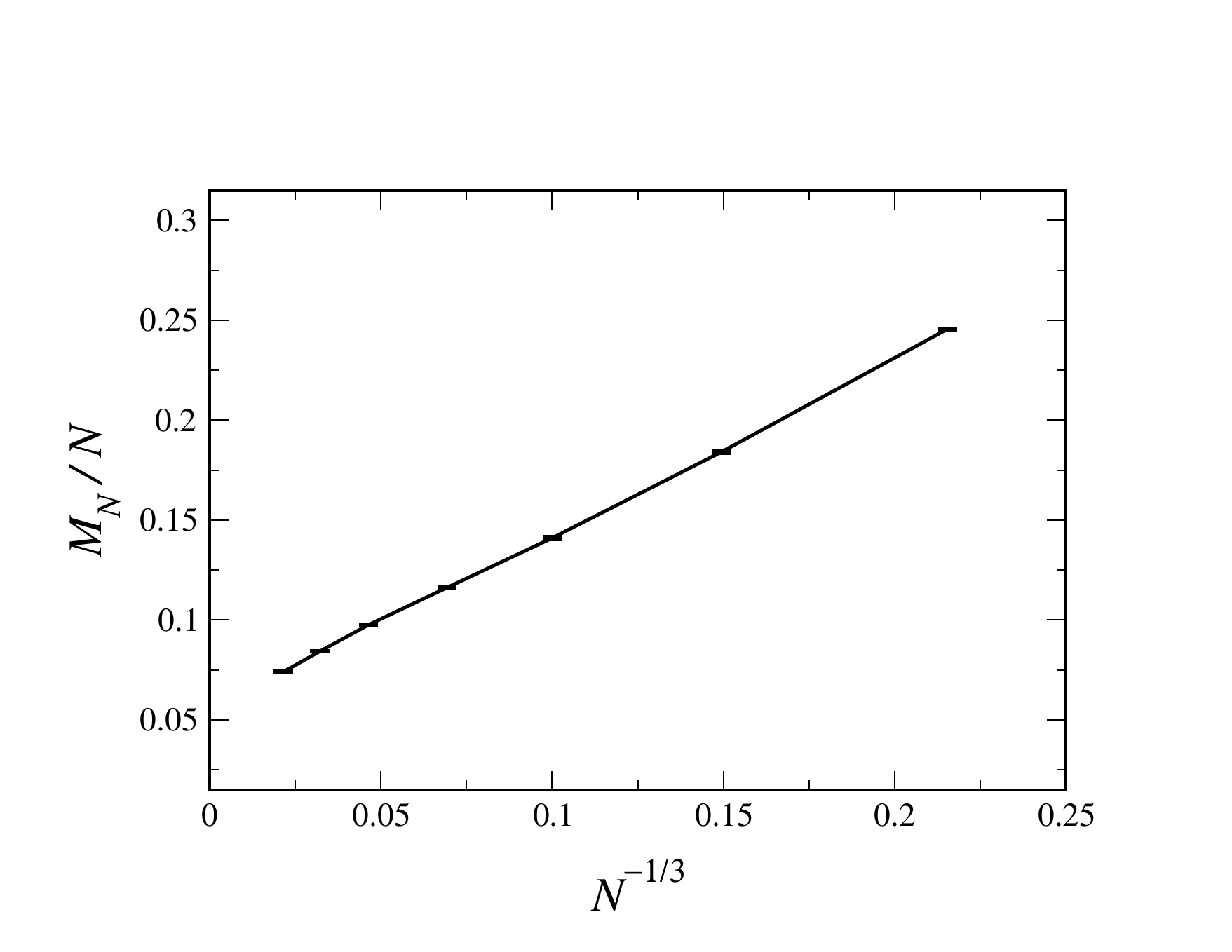}\\ 
\caption{The scaled ``mass" of the aggregate, $M_N/N$, for the SIS
model in two dimensions. The Ginzburg analysis suggests that, for
large $N$, this ratio approaches a constant. Indeed, the plot shows
that as $N$ increased  the ratio converges to a finite value.
However this convergence is very slow, as implied by the $N^{-1/3}$
scaling of the abscissa.}\label{fig2}
\end{figure}

For 2d SIS, our expression predicts $\kappa = 1$, so that the size of
the stochastic regime is of the same order as the self interaction
$1/N$ corrections. As shown above, $\kappa$ determines also the
relation between the average size of the epidemic and $N$ when the
infection rate takes its $N \to \infty$ critical value, $R_0 =1$.
Thus, in this case, we expect $M_N \sim N$. The data for this is
presented in Fig. \ref{fig2}. The results are indeed consistent with
the prediction; however the convergence is quite slow, much slower
than in 0 and 1 dimensions.

In three dimensions, $\kappa > 1$ and so the transition region is
smaller than the ${\cal{O}}(1/N)$ (self interaction) shift in the
transition point. This leads to an interesting situation where there
are two separate scaling regimes for large $N$.  We will return to
this point after first discussing the case of the SIR model.

\section{The Susceptible-Infected-Recovered (SIR)  model on spatial domains. }

The other classic model of epidemics is the SIR model, which assumes that a recovered
($R$) individual cannot be infected again, so it is removed
irreversibly from the "pool" of susceptible. The basic processes
are:
\begin{equation}
S+I \stackrel{\alpha/N}{\longrightarrow} 2I  \qquad I
\stackrel{\beta}{\longrightarrow} R.
\end{equation}
The corresponding master equation for the microscopic process in a
well-mixed population can be formulated in terms of $m$, the number
of susceptibles, and $n$, the number of infected individuals:
\begin{equation} \label{master}
\dot{P_n} = \beta \left( -nP_{m,n} + (n+1)P_{m,n+1} \right) +
\frac{\alpha}{N} \left(-nmP_{m,n} + (n-1)(m+1) P_{m+1,n-1} \right).
\end{equation}
In the deterministic limit, with the definition $S= \sum_m mP_{m,n}$
and $I = \sum_n n P_{n,m}$ and neglecting correlations ($\overline{nm} =
\bar{n} \bar{m}$) one gets the equations:
\begin{equation}
\dot{S} = -\frac{\alpha}{N} SI \qquad \dot{I} = -\beta I
+\frac{\alpha}{N} S I \qquad \dot{R} = \beta I,
\end{equation}
where the last equation is just a consequence of the $I$ dynamics.
Since $S = N-R-I$, the two coupled equations, (again introducing
$\widetilde\Delta = \alpha/\beta-1$):
\begin{equation} \label{qq}
\dot{I} =  \beta\widetilde\Delta I -\frac{\alpha}{N} I^2 - \frac{\alpha}{N} I R
\qquad \dot{R} = \beta I,
\end{equation}
are enough to describe the system. The SIR dynamics does not support
a  nontrivial equilibrium steady state; instead at any site the
epidemic disappears when $t \to \infty$, leaving a finite density of
recovered behind. This is manifested by the irreversible dynamics of
$R$.

Clearly, given $I(x,t)$ one can solve for the number of recovered
individuals at $x$:
\begin{equation}
R(x,t)  = \beta
\int_0^t I(x,\tau) d\tau,
\end{equation}
Plugging that into Eq. (\ref{qq}) and adding terms that represent
migration and discrete noise one gets:
\begin{equation} \label{qq1}
\dot{I} = D \nabla^2 I + \beta\widetilde\Delta I -\frac{\alpha}{N} I^2 -
\frac{\alpha\beta}{N} I \int_0^t I(\tau) d\tau + \eta(x,t) \sqrt{I}
\end{equation}
where $\eta$ is a delta-correlated noise, and $D=\alpha\chi/2$ is
the effective diffusion constant. Naive scaling analysis of
Eq. (\ref{qq1}) shows that the $I^2$ term is irrelevant and that the
noise term becomes relevant when $d \leq 6$, as expected from the
mapping to the dynamic percolation problem. Following
\cite{janssen-tauber} we integrate both sides of Eq. (\ref{qq1}) from
$t=0$ to $\infty$, using $\int \dot{I} dt = 0$ and $\int_0^\infty
I(t) \int_0^t I(\tau) dt d\tau = \nicefrac{1}{2} [\int_0^\infty I(t)
dt]^2$, we arrive at
\begin{equation} \label{qq2}
 D \nabla^2 \Phi   + \beta\widetilde\Delta  \Phi -\frac{\alpha\beta}{2N} \Phi^2  + \zeta(x)
 \sqrt{\Phi} = 0.
\end{equation}
where $\Phi(x) \equiv \int_0^\infty I(x,t) dt$. Note that the
variance of the noise term in Eq. (\ref{qq2}) must satisfy
\begin{equation}
\overline{\rm{Noise}^2} = \int_0^\infty dt_1 \int_0^\infty dt_2
\overline{\eta(t_1) \eta(t_2)} \sqrt{I(t_1)} \sqrt{I(t_2)} =
\int_0^\infty dt_1 I(t_1) = \Phi
\end{equation}
justifying the form of the noise amplitude term in the $\Phi$
equation. Rescaling Eq. (\ref{qq2}) by $N$ we have (now $m \equiv \beta
\widetilde\Delta$),
\begin{equation} \label{qq3}
 D \nabla^2 \Phi   + m  \Phi = \frac{\alpha\beta}{2} \Phi^2  + \frac{1}{\sqrt{N}} \zeta(x)
 \sqrt{\Phi}.
\end{equation}
Eq. (\ref{qq3}) may be analyzed perturbatively, as shown by
\cite{janssen-tauber}, by the same diagrammatic expansion used for
the directed-percolation case (see Appendix 1) where the only
difference is that the free propagator, instead of Eq. (\ref{prop}), is
\begin{equation} \label{prop1} \langle \Phi(k) \Phi(k')  \rangle =
\delta(k'+k)  \frac{1}{k^2+m}.
\end{equation}
The first correction to the diffusion constant comes from the same
self-energy diagram shown in Fig. \ref{diagram}, but here the
correction is $[N\widetilde\Delta^{\frac{6-d}{2}}]^{-1}$, thus $\kappa =
2/(6-d)$. Accordingly, in both the SIR and SIS cases, we have that
\begin{equation}
\kappa = \frac{2}{d_u - d},
\end{equation}
 where $d_u$ is the upper critical
dimension.

\begin{figure}
\vspace*{-.4in}\includegraphics[width=0.7\textwidth]{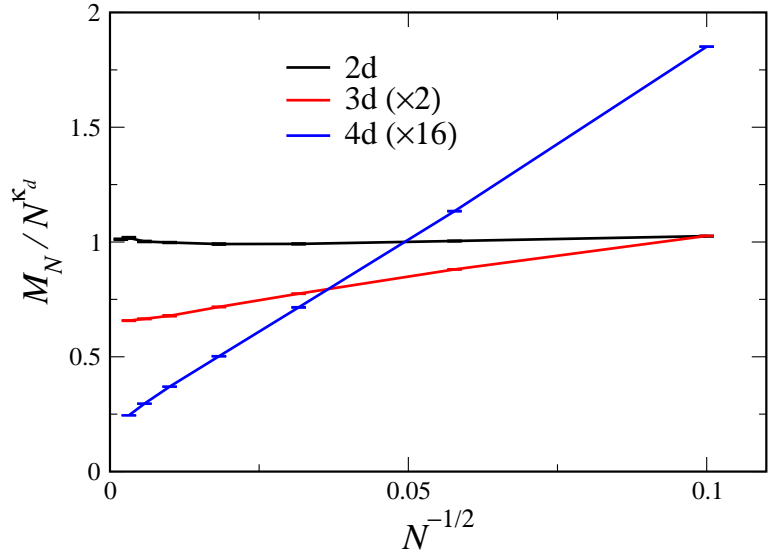}\\ 
\caption{The scaled ``mass" of the aggregate, $M_N/N$, for the SIR
model in $D=2$, $3$, and $4$, showing the convergence to a finite
value in the limit $N\to\infty$.} \label{fig4}
\end{figure}

This result is consistent with the exact scaling of the transition
region in SIR in 0 dimensions, namely $\kappa=1/3$ \cite{bennaim}.
It is also consistent with our numerical findings in I for the case
of one dimension, where we found $\kappa \approx 0.41$, to be
compared with our prediction of $2/5$. We can test our prediction
for higher dimensions by again measuring the total mass at the
classical transition point divided by $N^\kappa$.  This is presented
in Fig. \ref{fig4}. The results are seen to converge relatively
quickly to its finite $N=\infty$ value in two and three  dimensions,
but show, similar to the SIS case in two dimensions, a very slow
convergence in four dimensions.

\section{The  case of  $\kappa > 1$}

As we have seen above, two scales are involved in the large $N$
limit. One is the shift of the transition point due to the absence
of self-interactions, and this leads to $1/N$ corrections for the
critical reproductive number $R_c$, and the other is the width of
the ``quantum" regime where fluctuations dominate the system
behavior, the width of this region scales like $N^{-\kappa}$. For $d
< d_u -2$ we obtained $\kappa <1$ and the quantum regime is wider
than the self-interaction shift, thus the effect of
self-interactions is negligible. If $d = d_u-2$ both corrections
scale like $N^{-1}$ and this leads to the slow convergence of the
results to the large $N$ limit. We still have to consider the case
where $\kappa > 1$, i.e., where the quantum regime is narrower than
the self interaction shift.

For the SIS and SIR dynamics considered here, and for an integer
number of dimensions, we have to consider the case $\kappa = 2$ for $d=d_c-1$ (3d
for SIS, 5d for SIR) and $\kappa = \infty$ at the upper critical
dimension.

At $d=d_u$ the situation is trivial: $\kappa = \infty$ means that
the width of the transition zone is zero, since the system behaves
(up to logarithmic corrections) like its mean-field (infinite
dimensional) limit. Note the difference between a well-mixed (0d)
and the mean field ($\infty$d) cases: in the first there is a
pronounced quantum regime at finite $N$. In the second each point
has infinite number of neighbors so the ``effective $N$" is infinite
even if the number of individuals at each point is finite.

\begin{figure}
\vspace*{-.4in}\includegraphics[width=0.7\textwidth]{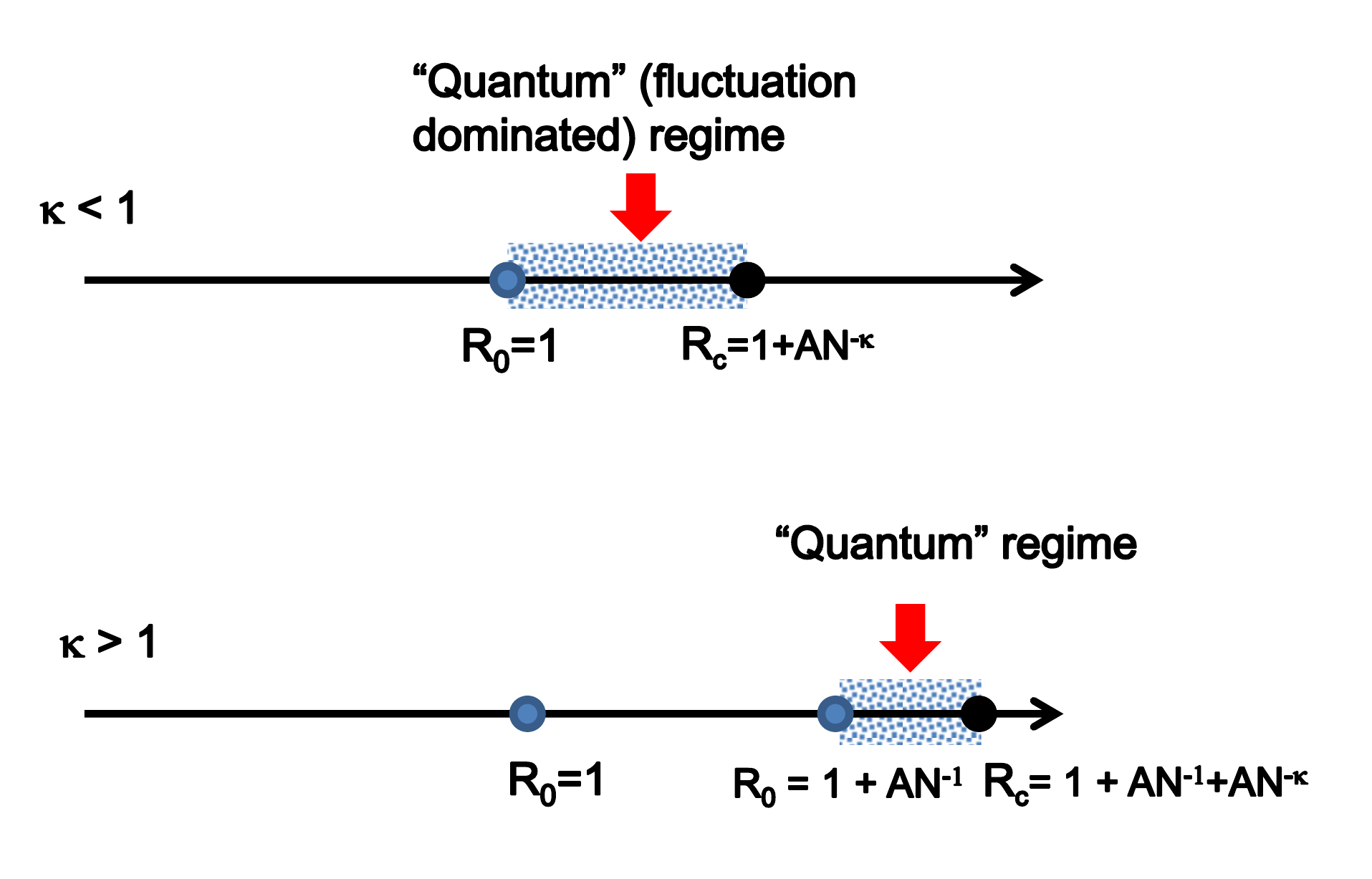}\\ 
\caption{The two possible scenarios for large $N$ scaling. If
$\kappa < 1$ (upper sketch) the $1/N$ self interaction  shift is
negligible with respect to the width of the quantum regime, thus the
convergence to the deterministic limit is controlled by a single
parameter $N^{-\kappa}$. The case $\kappa > 1$  (lower) is dominated
by two scales: one that controls the distance of the transition
point from its deterministic value, and the other that determines
the width of the fluctuation dominated zone.  } \label{kappa}
\end{figure}

What remains is   $d=d_u-1$, namely three dimensions for SIS and
five dimensions for SIR.  In these cases,  $\kappa=2$, so  the
``quantum" regime has a very small width (of order $N^{-2}$) around
the quantum transition point, which in turn is at a much larger
distance (of order $1/N$) away from the deterministic transition
point $R_0 = 1$. The situation is summarized in Fig. \ref{kappa}.

Although the transition point  converges to $R_0 =1$ in the infinite
$N$ limit, this convergence is slower than the rate in which the
quantum zone shrinks around this point. This gives rise to two
different scaling regimes, one of width $1/N$ and the second of
width $1/N^2$. We can see this behavior, again, by  studying $M_N$,
the total mass of the infection, now as a function of $R_0$, the
"bare" reproductive number.

In the outer region, with width ${\cal{O}}(1/N)$, the mass obeys the scaling law
\begin{equation}
M_N = N^{\tau_M^{\textit{out}}} {\cal G}_\textit{out}\left(\widetilde\Delta N\right).
\end{equation}
Now, far from the transition point (say for fixed $R_0$ slightly below 1), at large enough $N$ the dependence of $M_N$ on
this distance must approach its MF limit, $M_N \sim 1/(1 - R_0)$, independent of $N$. This implies that for large negative argument, ${\cal{G}}(x) \sim -1/x$, and that $\tau_M^\textit{out} = 1$.  Since the transition point is at
$R_{c} \approx 1 +  (A/N) +  (B/N^2)$, where $A$ and
$B$  are some constants, $M_N$ must get large as $\widetilde \Delta N$ approaches $A$.
Since in this outer region, fluctuations are small, the incipient divergence of $M$ is mean-field like, so ${\cal{G}}_\textit{out}$ diverges as ${\cal{G}}_\textit{out} \sim c/(A-x)$, so that for $R_0$ near, but not too near $R_c$, $M_N$ behaves as
\begin{equation}
M_N \approx \frac{C}{1 + A/N - R_0}
\end{equation}

\begin{figure}
\includegraphics[width=0.7\textwidth]{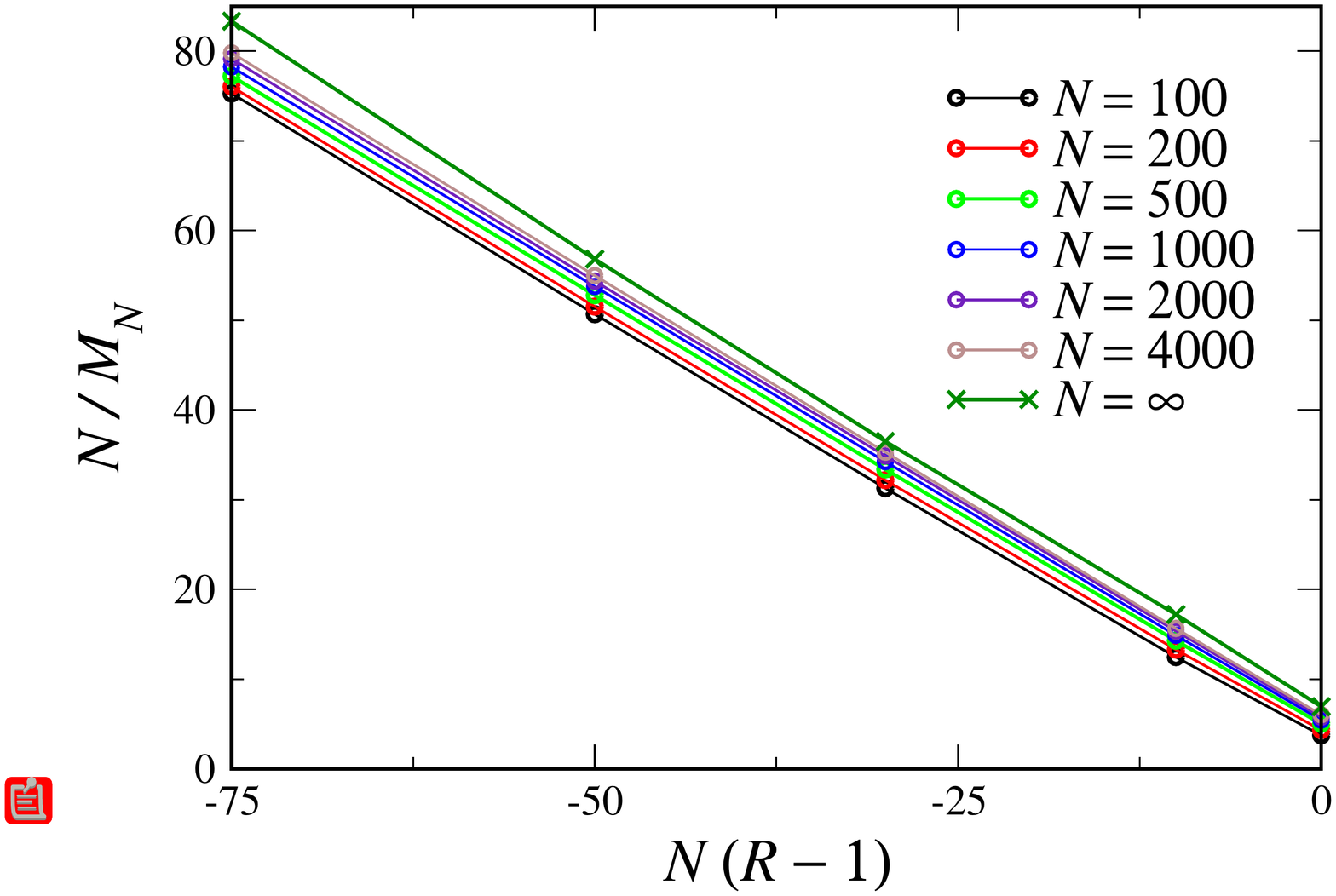}\\
\includegraphics[width=0.7\textwidth]{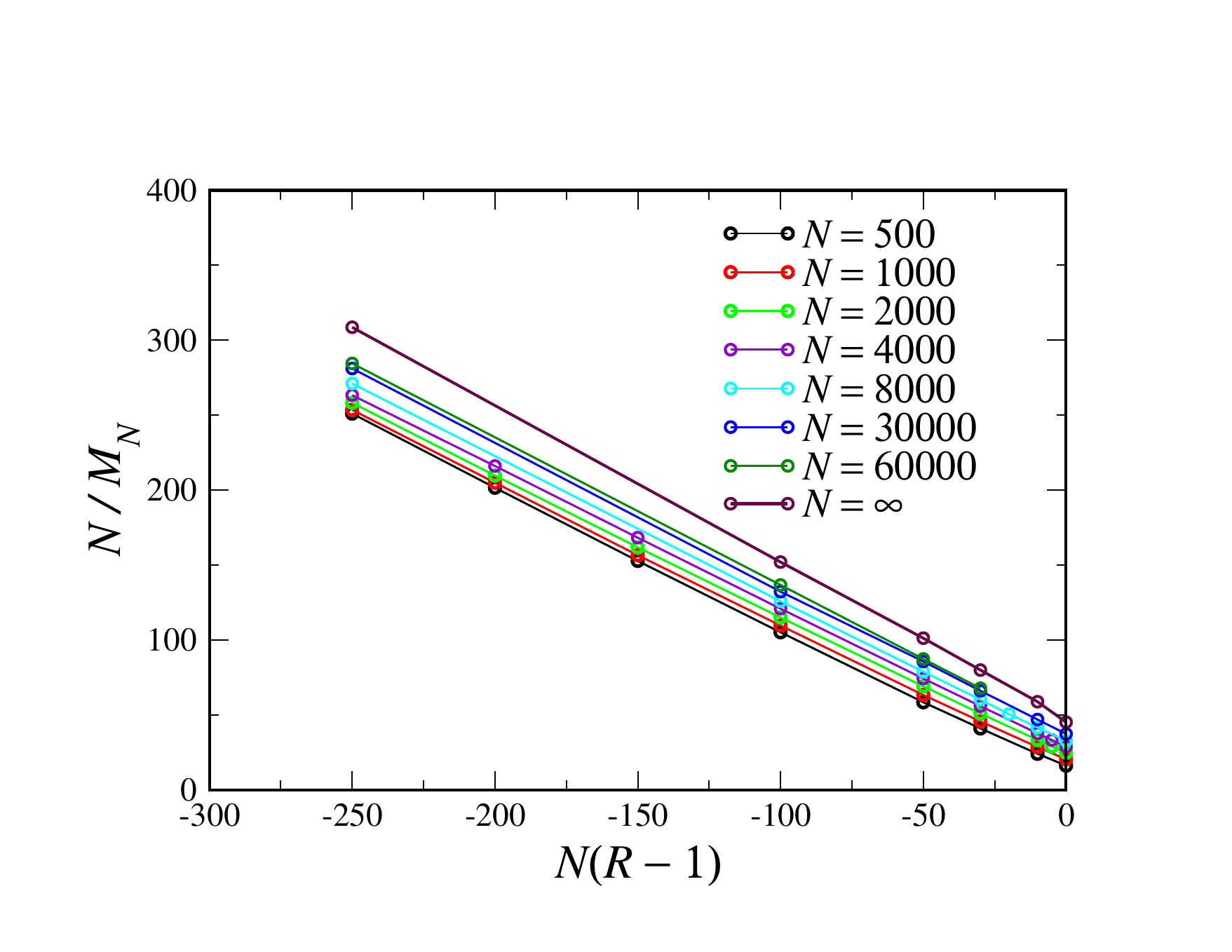}
\caption{Upper Panel:  The inverse of the scaled ``mass'' of the
aggregate, $N/M_N$, for the SIS model in $d=3$ as a function of
$N(R_0-1)$, for various $N$.  The behavior for large $N(R_0-1)$ is
consistent with $M_N =1/(1-R_0)$.  The data labelled $N=\infty$ was
obtained by fitting a quadratic curve in $N^{-1/3}$ to $M_N$ for
fixed $N(R_0-1)$ and extrapolating.  This $N=\infty$ curve fits well
to $M_N=N/(7.2-N(R_0-1))$, corresponding to a shift in the critical
$R_0$ by an amount $7.2/N$.  Lower Panel: The inverse of scaled
``mass'' of the aggregate, $N/M_N$, for the SIR model in $D=5$ as a
function of $N(R_0-1)$, for various $N$.  The behavior for large
$N(R_0-1)$ is consistent with $M_N =1/(1-R_0)$.  The data labelled
$N=\infty$ was obtained by fitting a quadratic curve in $N^{-1/3}$
to $M_N$ for fixed $N(R_0-1)$ and extrapolating.  This $N=\infty$
curve fits well to $M_N=N/(47.6-N(R_0-1))$, corresponding to a shift
in the critical $R_0$ by an amount $47.6/N$.} \label{outer}
\end{figure}

This behavior is demonstrated in  Fig. \ref{outer}, where
$N/M_N$ is plotted  versus $\widetilde \Delta N$, for SIS in three dimensions
in the upper panel and for SIR in five dimensions in the lower.  We
see that there is a very slow convergence to an asymptotic curve.
This slow convergence to the asymptotic scaling limit is reminiscent
to what we encountered in the case of $d=d_u-2$. The large $N$ line
is straight, but does not converge to zero at $R_0=1$, since the
actual transition happens at $R_0 \approx 1 + A/N$. Although at
large $N$, the distance of $R_0=1$ from the transition shrinks to
zero one observes no "quantum" effects in the outer region since the
width of the quantum regime shrinks even faster.

As we approach very close,  of order a small fraction of  $1/N^2$, to  the phase transition point, the fluctuations become significant and $M_N$ diverges as
\begin{equation}
M_N = A N^{-\tau_m} (R_{c}-R_0)^{-\gamma}
\end{equation}
where $\gamma$ is the scaling exponent for the mass, which for DP in
three dimensions is $\gamma\approx 1.24$~\cite{Jensen} and  is
approximately $1.2$ for percolation in five dimensions. The general scaling law in the inner region, of width ${\cal{O}}(1/N^2)$ is then
\begin{equation}
M_N = N^{\tau_M^\textit{in}}  \left[{\cal G}_\textit{in}\left(\Delta N^2\right)\right]^{-\gamma}.
\end{equation}
where ${\cal{G}}_\textit{in}(x)$ vanishes linearly at $x=0$.  For large negative argument, this has to match onto the outer behavior for $\widetilde \Delta N \ll 1$.
This is possible if ${\cal{G}}_\textit{in}(x) \sim -Cx$ as $x \to \infty$ and $\tau_M^\textit{in} = 2$.

\begin{figure}
\includegraphics[width=0.7\textwidth]{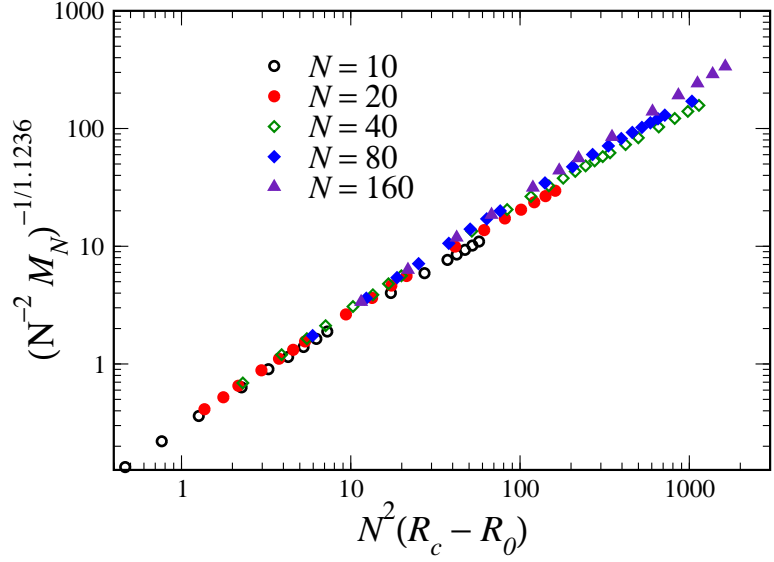}//
\includegraphics[width=0.7\textwidth]{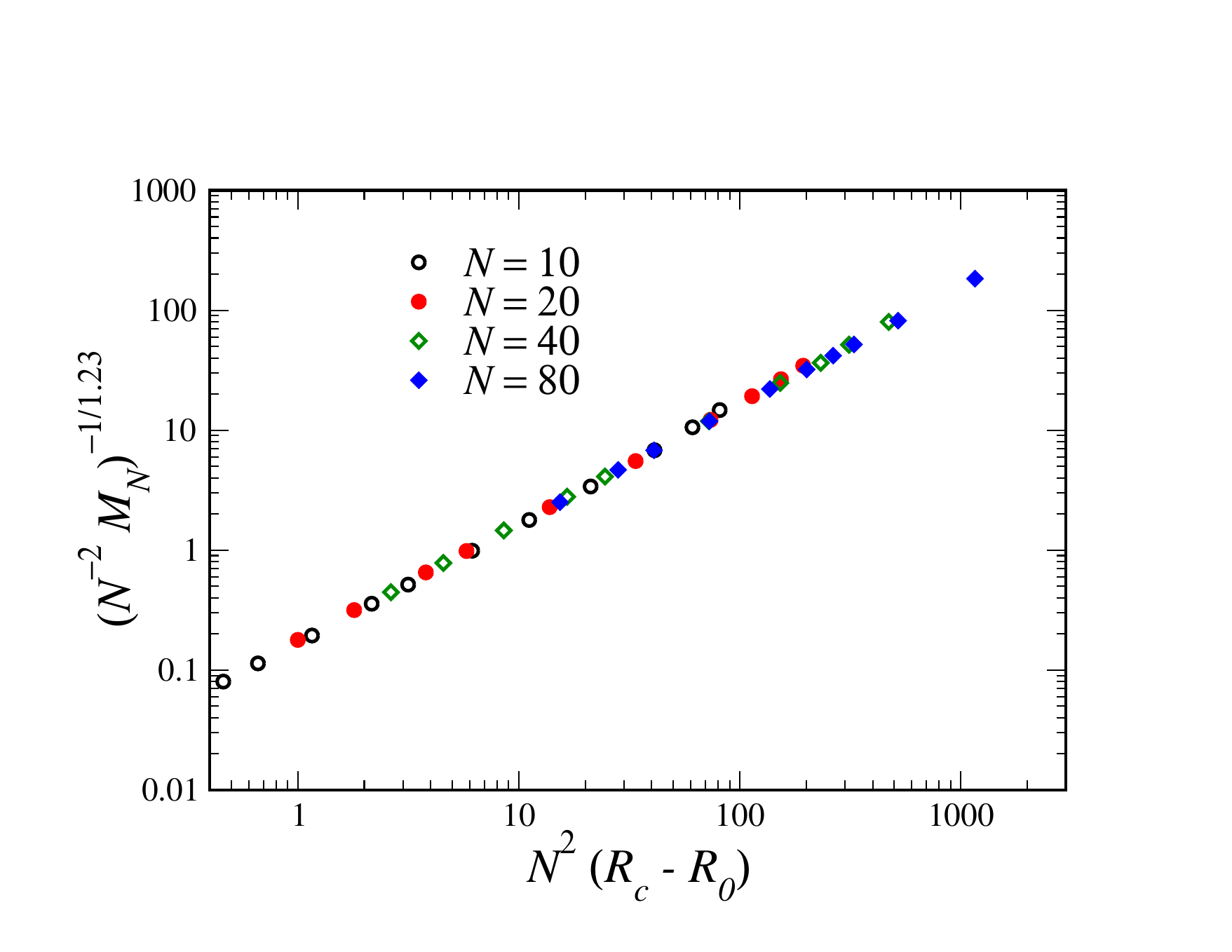}
\caption{(color online) Top: The scaled ``mass'' of the aggregate,
$(M_N/N^2)^{-1/\gamma}$, as a function of the scaled ``inner''
variable $N^2(R_c - R_0)$, for the SIS model in $D=3$.  We used the
value $\gamma=1.236$.  Bottom: The scaled ``mass'' of the aggregate,
$(M_N/N^2)^{-1/\gamma}$, as a function of the scaled ``inner''
variable $N^2(R_c - R_0)$, for the SIR model in $D=5$.  We used the
value $\gamma=1.23$.} \label{inner}
\end{figure}

 Accordingly, the plot of
$(N^{-2} M_N)^{-1/\gamma}$ vs. $\Delta N^2$ shows the inner scaling
function ${\cal{G}}_\textit{in}$ in the large $N$ limit and goes linearly to zero at
the transition point. This behavior is demonstrated in Fig.
\ref{inner} for both the 3d SIS (upper panel) and the 5d SIR (lower
panel) models.

\section{Front Velocity}
In the wake of Brunet and Derrida's \cite{derrida} pathbreaking work
on the large $N$ behavior of the front velocity in Fisher-type
systems, there has been an enormous amount of attention devoted to
this issue, including a rigorous proof of the original heuristic
arguments.  It is thus natural to ask how this work relates to our
current findings.  The first thing to note is that the limits
addressed here and the result of \cite{derrida} are different. The
Brunet-Derrida limit corresponds in our language to fixed $\Delta$,
$N\to \infty$, whereas we are interesting in the limit $\Delta \ll
1$, $N \gg 1$, $\Delta N^\kappa \sim {\cal{O}}(1)$.

We first investigate the behavior in the immediate vicinity of the transition point, restricting our attention to the 1d SIS model.  In the immediate vicinity of the
transition point, both the spatial correlation length, $\xi_\perp$, and the time correlation scale, $\xi_\parallel$ diverge. It is expected then that the velocity will scale, in this regime, as the ratio of $\xi_\perp$ to $\xi_\parallel$:
\begin{equation}
v \approx \frac{B_\perp N^{-\kappa(\nu_\perp - \frac{1}{2})}
\Delta^{-\nu_\perp}}{B_\parallel
N^{-\kappa(\nu_\parallel-1)}\Delta^{-\nu_\parallel}} = B_v
N^{\kappa(\nu_\parallel - \nu_\perp - \frac{1}{2})}
\Delta^{\nu_\parallel-\nu_\perp} =B_v N^{-\kappa/2} \left(\Delta
N^\kappa \right)^{\nu_\parallel-\nu_\perp}
\end{equation}
Since $\nu_\parallel > \nu_\perp$, the velocity
vanishes as the
transition point is neared, just as in the classical theory.
Furthermore, $\nu_\parallel -\nu_\perp - 1/2>0$, so the velocity
increases with $N$ for fixed $\Delta$.  This is consistent with the
Brunet-Derrida asymptotic result, which also has the velocity rising
with $N$ at fixed $\Delta$.

\begin{figure}
 \includegraphics[width=0.7\textwidth]{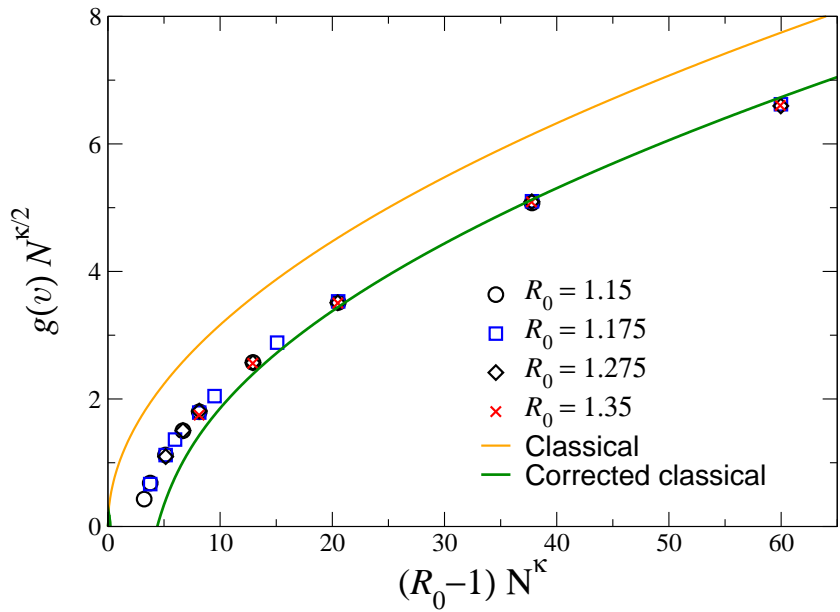}
 \caption{The scaling collapse of the scaled transformed velocity,
 $g(v) N^{\kappa/2}$, where $g(v)$ is given in Eq. (\ref{gdef}),
 versus $\widetilde\Delta N^\kappa$ for the 1d SIS model, with $\beta=1$, $\chi = 0.2$.
 The ``classical" result is $\sqrt{(R_0-1)N^\kappa}$, whereas the
``corrected classical" result is
 given by Eq. (\ref{bd}).}
 \label{figv}
 \end{figure}

In the classical limit the front velocity is given by  $v \sim
\sqrt{\widetilde\Delta}$, independent of $N$.  One is tempted, then,
to write, in analogy with our other scaling laws, $v \approx
N^{-\kappa/2} {\cal{H}}(\widetilde \Delta N^{\kappa})$.  The problem
with this is that, while in the continuum classical limit, the
velocity is proportional to $\sqrt{\widetilde\Delta}$, on the lattice this is
true only for small $\widetilde\Delta$. To work with discrete agents and to
define their local density one should implement some UV cutoff, so
even for off-lattice models the relevant result is the one obtained
for a lattice.  The classical lattice velocity $v_L$ satisfies the
equation (see \cite{pat})
\begin{equation}
 \frac{v_L}{\alpha \chi} \ln\left(\frac{v_L}{\alpha \chi} +
 \sqrt{1 + \left( \frac{v_L}{\alpha\chi}\right)^2}\right) + 1 -
\sqrt{1 + \left(\frac{v_L}{\alpha\chi}\right)^2} =
\frac{\beta\widetilde\Delta}{\alpha\chi}
 \end{equation}
 so that, for large $\widetilde\Delta$, the
 velocity grows as $\widetilde\Delta/\ln(\widetilde\Delta)$, as
opposed to $\sqrt{\widetilde\Delta}$.  Thus, instead of trying to
find a scaling relation for $v$, it is preferable to find a scaling
relation for
 \begin{equation}
 g(v) \equiv \sqrt{ \frac{\alpha\chi}{\beta}}
 \left[  \frac{v}{\alpha \chi} \ln\left(\frac{v}{\alpha \chi} + \sqrt{1 +
 \left( \frac{v}{\alpha\chi}\right)^2}\right) + 1 -
 \sqrt{1 + \left(\frac{v}{\alpha\chi}\right)^2}\right] ^{1/2}
 \label{gdef}
 \end{equation}
 which, for $v=v_L$, is precisely
 equal to $\sqrt{\widetilde\Delta}$. In Fig. \ref{figv}, we show the
scaling collapse of $g(v) N^{\kappa/2}$ versus $\widetilde\Delta
N^\kappa$.  The Brunet-Derrida effect, namely the anomalously slow
approach to the classical velocity, is apparent from this graph,
where even for $\Delta N^{-\kappa}\sim 60$, the scaling curve is
very far below  the classical result.

In more detail, for large positive argument, the Brunet-Derrida
result implies that
 \begin{equation}
 {\cal{H}}(x) \approx \sqrt{x} \left ( 1 - \frac{9\pi^2}{4\ln^2 x}\right)
 \label{bd}
 \end{equation}
 This corrected classical result is also show in Fig. \ref{figv}, where
we see quite good agreement, especially considering the relatively
small values of $N$ involved, compared to those necessary to achieve
even semi-quantitative agreement with the Brunet-Derrida correction
at $\widetilde\Delta \sim {\cal{O}}(1)$.

\section{ Summary and Discussion}

Along this paper we have studied, numerically and analytically, the
convergence of the stochastic process to the deterministic rate
equations when the number of particles is large. In spatially
extended model there are two parameters that control the
convergence: the number of particles per site $N$ and the distance
from the transition point. Together, these parameters yield a region
of size $\Delta N^{-\kappa}$ above the phase transition point;
within this region the system is dominated by demographic noise and
the deterministic equations fail to describe it accurately.

The value of $\kappa$ has been found before using an extensive
analysis of zero dimensional \cite{bennaim, kess, ks1, lof} and one
dimensional \cite{KS0} models. It turns out that this particular
result may be derived directly, for any dimensionality, using the
Ginzburg analysis. For the fundamental models considered here it
depends only on the difference between $d_u$, the upper critical
dimension, and $d$, via $\kappa =  2/(d_u-d)$.

Clearly, this general expression stems from the fact that the
leading perturbative correction (i.e., the diagrams that lead to  an
infrared divergence in the highest dimension, which is thus the
upper critical dimension) is proportional to $1/N$, since it
involves an average over two noise terms, each is proportional to
$1/\sqrt{N}$. This seems to be a generic property of stochastic
processes and will be interesting to find out a model for which this
general argument is not applicable.

Below $d_u-2$ $\kappa <1$, and the self-interaction shift is
negligible at large $N$. In this case the point $R_0=1$ is peculiar:
its normalized distance from the critical point (the distance
divided by the width of the quantum regime) is $N$ independent.
Accordingly, the divergence of various observables at this point is
determined solely by $N^{\kappa}$. This feature facilitates the
numerics, since one can extract the value of the exponent without
finding $R_c$. If $\kappa >1$ this is no longer the case, and to
locate the quantum regime one has to first identify the transition
point.

Although the SIS and SIR processes serves us here as an archetypic
stochastic processes that belong to the most pronounced equivalence
classes of out-of-equilibrium transitions, they are also interesting
models for epidemiologists. Several attempts have been made,
recently, in order to understand better the role of fluctuations in
individual-based, spatially structured epidemic models. The results
presented here practically solve this problem for the case of
subpopulations on a lattice considered in \cite{getz}.

In the common case of zoonotic infections the pathogens first
emerged from animal reservoirs, inducing a "stuttering transmission"
stage in which $R<R_c$, and reaching the phase of sustained
transmission (human outbreak) only due to pathogen evolution (in
human environment) to $R_0 > R_c$ \cite{zoo1}. If $R_0$ is growing
slowly to larger value (as opposed to a major evolutionary step
caused by a single mutation) the pathogen must cross the quantum
region, where the size of the outbreak (the number of infections,
and hance the chance for the next evolutionary step to occur) is
simply $M_N(R)$. With an appropriate knowledge about the adaptation
process of the pathogen, it will be quite easy to implement our
results to obtain the chance of an outbreak.

\section{Acknowledgement}  The authors thank Prof. Pierre Hohenberg
for pointing out the work of Mon and Binder on the MF limit of
equilibrium systems.

\section{Appendix 1}

 Here we show some elements of the perturbative expansion of the action
(\ref{action}, \ref{S1}) and the terms that determine the leading
correction for large $N$, as explained in the text.

\begin{figure}
\vspace*{-.4in}\includegraphics[width=0.7\textwidth]{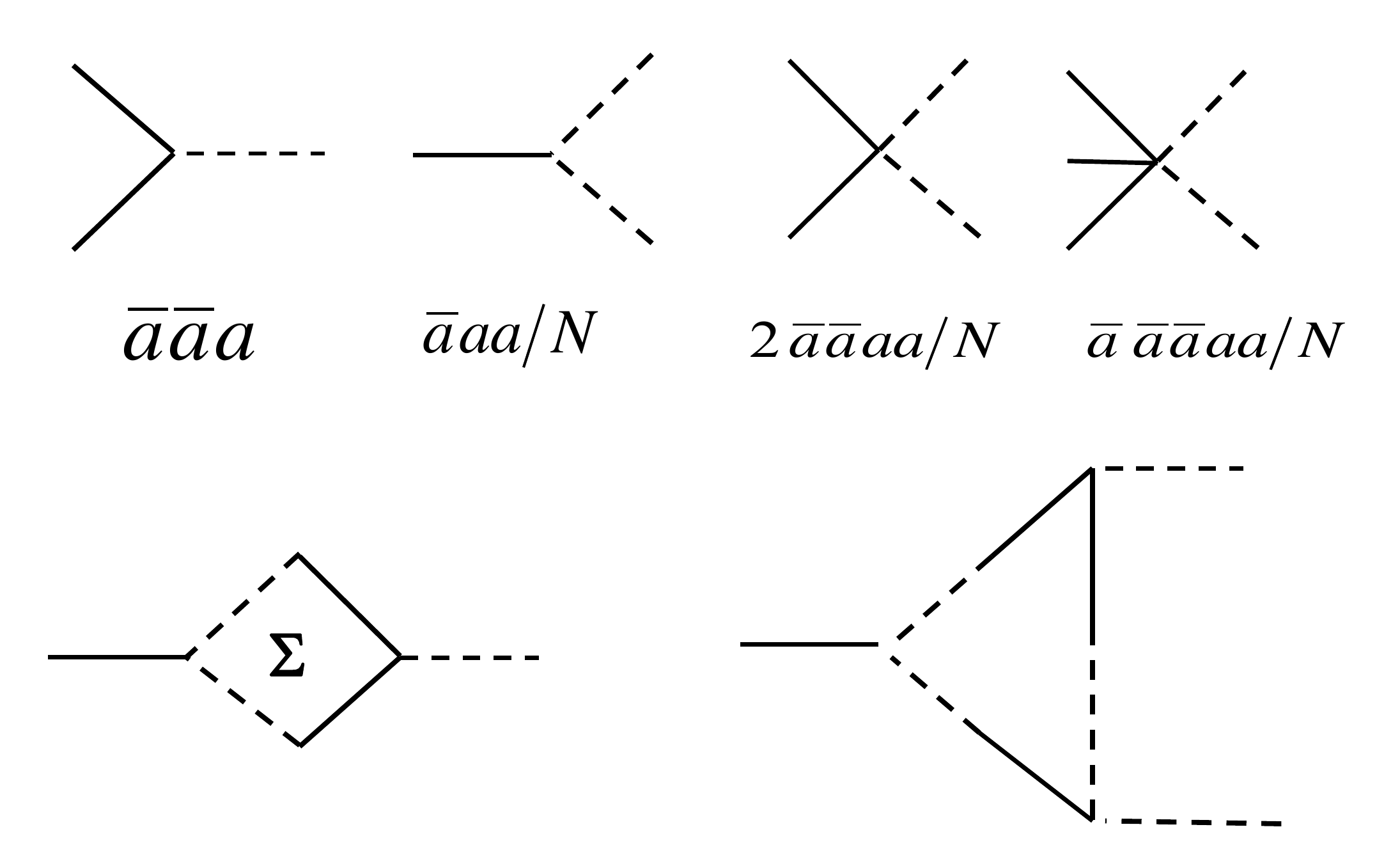}\\ 
\caption{Element of the diagrammatic perturbative expansion. The
terms that appear in Eq. (\ref{S1}) (upper line), the self energy
diagram $\Sigma$ and the 1-loop correction to the three point vertex
(lower part).  } \label{diagram}
\end{figure}

The elementary diagrams that appear in the perturbative expansion
are shown in the upper part of Fig. \ref{diagram}. Of those, the
first two appear in the Reggeon field theory  and  yield the
one-loop renormalization of the mass and $R_0$. The  diagrams
involved are presented in the lower part of Fig. \ref{diagram}.

With the bare propagator, Eq. (\ref{prop}), one can see that the leading
correction to the mass behaves like $$\frac{1}{N}\int
\frac{q^{d-1}dq}{q^2+m}.$$ This implies that $q$ scales like
$\sqrt{m}$ and hence close to the transition the result is
proportional to $\Delta^{-(2-d)/2}$, thus from this diagram one
would get $\kappa =2/(2-d)$ ($x=(2-d)/2, \ y=1$, see text). The
triangular diagram that provides the correction to the coupling
constant scales like $$\frac{1}{N^2}\int
\frac{q^{d-1}dq}{(q^2+m)^2},$$ so it corresponds to $\kappa =
4/(4-d)$. However, the corrections to the diffusion constant are
given by the second derivative of the self-energy diagram with
respect to the incoming momentum, and this contribution is
proportional to $$\frac{1}{N}\int \frac{q^{d-1}dq}{(q^2+m)^2},$$ and
this term yields the minimum value $\kappa = 2/(4-d)$ given in Eq.
(\ref{siskappa}).


\begin{thebibliography}{S}




\bibitem{vanKampen}N. G. van Kampen, \emph{Stochastic Processes in Physics and Chemistry} (North-Holland, Amsterdam,
1992).
\bibitem{extinct}See, e.g., D. A. Kessler and N. M. Shnerb, J. Stat. Phys. \textbf{127}, 861 (2007); O. Ovaskainen
and B. Meerson, Trends Ecol. Evol. \textbf{25}, 643 (2010).
\bibitem{derrida}E. Brunet and B. Derrida, \pre \textbf{56}, 2597 (1997).
\bibitem{nks}D. A. Kessler, Z. Ner and L. M. Sander, \pre \textbf{58}, 107 (1998).
\bibitem{KS0}D. A. Kessler and N. M. Shnerb, J. Phys. A: Mathematical and Theoretical \textbf{41}, 292003 (2008).
\bibitem{sis}G. H. Weiss and M. Dishon, Math. Biosci. \textbf{11}, 261 (1971).
\bibitem{hinrichsen}H. Hinrichsen, Adv. Phys. \textbf{49}, 815 (2000).
\bibitem{pelissetto} A. Pelissetto, P. Rossi and E. Vicari, \pre \textbf{58}, 7146 (1998).
\bibitem{binder}K. K. Mon and K. Binder, \pre \textbf{48},  2498 (1993).
\bibitem{sir}W. O. Kermack and A. G. McKendrick, Proc. Roy. Soc. A \textbf{115}, 700 (1927).
\bibitem{anderson} R.M. Anderson and R.M. May, \emph{Infectious Diseases in Humans}, Oxford University Press, Oxford (1992).
\bibitem{kess}D. A. Kessler, J. Appl. Prob. \textbf{45}, 757 (2008).
\bibitem{durrett} R. Durrett and S.A. Levin,  Theor. Pop. Biol. \textbf{46}, 361 (1994).
\bibitem{yosibz}Y. Ben-Zion, Y. Cohen and N. M. Shnerb, J. Theor. Biol. \textbf{264}, 197 (2010).
\bibitem{fisher} R. A. Fisher, Ann. Eugenics 7, 353 (1937); A. N. Kolmogorov, I. G. Petrovskii and N. S. Piskunov, Selected
Works of A. N. Kolmogorov. V. M. Tikhomirov (Ed.), Kluwer Academic
Publishers, 1991.
\bibitem{pelity} M. Doi, J. Phys. A 9, 1465 (1976); L. Peliti, J. Physique 46, 1469
(1985).
\bibitem{cardy} J.L. Cardy and U.C. Ta\"{u}ber, J. Stat. Phys. \textbf{90}, 1 (1998).
\bibitem{janssen-tauber} H-K Janssen and U.C. T\"{a}uber, Annals of
Physics \textbf{315}, 147 (2005).
\bibitem{bennaim} E. Ben-Naim and P. L. Krapivsky, Phys. Rev. \textbf{E69} 050901(R) (2004).
\bibitem{Jensen}I. Jensen, \pra {45}, R563 (1992).
\bibitem{ks1} D.A. Kessler and N.M. Shnerb, Phys. Rev. \textbf{E. 76}
010901 (2007).
\bibitem{pat} L. Pechenik and H. Levine, Phys. Rev. \textbf{E 59}, 3893 (1999).
\bibitem{lof} A. Martin-L\"{o}f, J. Appl. Probab. \textbf{35}, 671 1998.
\bibitem{getz} W. M. Getz, et al., in Z. Feng, U. Dieckmann and S. Levin,
eds., \emph{Disease Evolution: Models, Concepts and Data Analyses},
DIMACS Series in Discrete Mathematics and Theoretical Computer
Science 71, (American Mathematical Society, Providence, RI, 2006),
p. 113.
\bibitem{zoo1} R. Antia et. al., Nature \textbf{426} 658
(2003); J.O. Lloyd-Smith, et al., Science 326, 1362 (2009).

















\end{thebibliography}
\end{document}